\documentclass[journal,twocolumn,a4paper]{IEEEtran}

\usepackage{amsmath}
\usepackage{amssymb}
\usepackage{cite}
\usepackage{graphicx}
\usepackage{subfigure}
\usepackage{multirow} 
\usepackage{longtable}
\usepackage{threeparttable}
\usepackage{caption3}
\usepackage{array}
\usepackage{cases}

\newtheorem{observation}{Observation}
\newtheorem{assumption}{Assumption}
\begin{document}

% title

\title{\LARGE{A Practical Cooperative Multicell MIMO-OFDMA Network Based on Rank Coordination}}

% authors
\author{Bruno Clerckx, Heunchul Lee, Young-Jun Hong and Gil Kim
\thanks{Manuscript Draft: \today.}%
\thanks{Bruno Clerckx is with Imperial College London, London SW7 2AZ, United Kingdom~(email: b.clerckx@imperial.ac.uk). Heunchul Lee and Gil Kim are with Samsung Electronics, Suwon-si, Gyeonggi-do 443-742, Republic of Korea~(email: heunchul.lee@samsung.com,giil.kim@samsung.com). Young-Jun Hong is with Samsung Electronics Co., Ltd., Samsung Advanced Institute of Technology, Yongin-si, Gyeonggi-do 446-712, Republic of Korea~(email: yjhong@ieee.org).}
\thanks{This paper was presented in part at IEEE Global Communications Conf. (Globecom), Dec. 2011, Houston, USA.}
}
\maketitle

\begin{abstract}
An important challenge of wireless networks is to boost the cell edge performance and enable multi-stream transmissions to cell edge users. Interference mitigation techniques relying on multiple antennas and coordination among cells are nowadays heavily studied in the literature. Typical strategies in OFDMA networks include coordinated scheduling, beamforming and power control. In this paper, we propose a novel and practical type of coordination for OFDMA downlink networks relying on multiple antennas at the transmitter and the receiver. The transmission ranks, i.e.\ the number of transmitted streams, and the user scheduling in all cells are jointly optimized in order to maximize a network utility function accounting for fairness among users. A distributed coordinated scheduler motivated by an interference pricing mechanism and relying on a master-slave architecture is introduced. The proposed scheme is operated based on the user report of a recommended rank for the interfering cells accounting for the receiver interference suppression capability. It incurs a very low feedback and backhaul overhead and enables efficient link adaptation. It is moreover robust to channel measurement errors and applicable to both open-loop and closed-loop MIMO operations. A 20\% cell edge performance gain over uncoordinated LTE-A system is shown through system level simulations.  
\end{abstract}


\section{Introduction}
\PARstart{I}{n} current wireless networks, the cell edge users experience low Signal to Interference and Noise Ratio (SINR) due to the high Inter-Cell Ifnterference (ICI) and cannot fully benefit from Multiple-Input Multiple-Output (MIMO) multi-stream transmission capability. Advanced interference mitigation techniques relying on multi-cell cooperation have drawn a lot of attention recently in the industry \cite{3GPPTR36.819} and academia \cite{Gesbert:2010}. Such techniques, commonly denoted as Coordinated Multi-Point transmission and reception (CoMP) in 3GPP LTE-A \cite{3GPPTR36.819}, are classified into joint processing (relying on data sharing among cells) and coordinated scheduling/beamforming (requiring no data sharing among cells).

\par This paper focuses on the second category requiring no data sharing. Three kinds of multi-cell cooperation are typically investigated, namely coordinated beamforming \cite{Dahrouj:2010,Cadambe:2008}, coordinated scheduling \cite{Choi:2008,Kiani:2008} and coordinated power control \cite{Gjendemsjoe:2008,Huang:2006}. Such cooperation types can be performed independently or be combined \cite{Venturino:2010,Venturino:2009,Yu:2010,Yu:2011}.

\par Despite the potential merits of such techniques in an ideal environment, it is shown in \cite{3GPPTR36.819,Gorokhov:2010} and confirmed in this paper that the benefits may vanish quickly in more practical scenarios due for instance to the fast variation of the inter-cell interference and inaccurate link adaptation, the sensitivity to Channel State Information (CSI) measurement, the quantized CSI feedback inaccuracy at the subband level, the limited payload size of the uplink control channels and the latency of the feedback and the backhaul. Unfortunately all those issues are most of the time neglected in the literature when it comes to the design and evaluations of multi-cell cooperative schemes. Indeed, it is assumed in \cite{Dahrouj:2010,Cadambe:2008,Choi:2008,Kiani:2008,Gjendemsjoe:2008,Huang:2006,Venturino:2010,Venturino:2009,Yu:2010,Yu:2011} that any local CSI can be available at the base station (BS) with no delay, no measurement error, no constraint on the uplink and backhaul overhead, no dynamic interference, with perfect CSI feedback on every subcarrier and with perfect link adaptation. Moreover, the receiver implementation is assumed perfectly known at the BS. 

\par Unlike previous papers that targeted optimal designs under ideal assumptions, other papers have focused on enhancing cooperative multi-cell schemes under non-ideal assumptions. In \cite{Papadogiannis:2008}, clustering is used to decrease the feedback overhead and reduce the scheduler complexity and the number of cooperating cells while conserving as much as possible the performance. In \cite{Tajer:2011}, the transmit beamformer is designed to account for imperfect CSI, modeled as noisy channel estimates. In \cite{Lee:2011}, limited feedback is considered and the feedback bits are allocated among cells in order to minimize the performance degradation caused by the quantization error. In \cite{Huang:2011}, an iterative algorithm is designed to optimize the downlink beamforming and power allocation in time-division-duplex (TDD) systems under limited backhaul consumption.

\par This paper provides a novel and practical multi-cell cooperative scheme relying on a joint user scheduling and rank coordination such that the transmission ranks (i.e.\ the number of transmitted streams) are coordinated among cells to maximize a network utility function. Theoretically, such a cooperative scheme is a sub-problem of the more general problem of a joint coordinated scheduling, beamforming and power control where the BSs control the transmission ranks by optimizing an ON/OFF power allocation on each beamforming direction. We could therefore adopt an iterative scheduler similar to the one used in \cite{Yu:2010,Yu:2011}. However, this paper aims at deriving a much simpler and practical scheme that directly addresses the problem of user scheduling and rank coordination without requiring the heavy machinery of the iterative scheduler. 

\par Unlike the referred papers \cite{Dahrouj:2010,Cadambe:2008,Choi:2008,Kiani:2008,Gjendemsjoe:2008,Huang:2006,Venturino:2010,Venturino:2009,Yu:2010,Yu:2011,Papadogiannis:2008,Tajer:2011,Lee:2011,Huang:2011} that account for at most one specific impairment, the cooperative scheme aims to be practical at the system level by accounting for impairments originating from both the terminal and the network constraints.

\par \textit{At the terminal side}, the rank coordination scheme relies on the report from the user terminal of a preferred interference rank, referring to the transmission rank in the interfering cell that maximizes the victim users' throughput, and a differential Channel Quality Indicator (CQI): 
\begin{itemize}
\item Such a report is of an implicit feedback type \cite{Clerckx:2010} and incurs a very low feedback overhead. An additional 2 bit feedback over uncoordinated LTE-A system is shown to bring 20\% cell-edge performance gain. Moreover the interference rank is a wideband information making it less sensitive to CSI measurement error.
This is in contrast with the explicit feedback of full and ideally measured CSI (i.e.\ the channel matrices between all users and their serving and interfering cells) on every subcarrier commonly assumed in the aforementioned approaches, e.g.\ \cite{Dahrouj:2010,Cadambe:2008,Choi:2008,Kiani:2008,Gjendemsjoe:2008,Huang:2006,Venturino:2010,Venturino:2009,Yu:2010,Yu:2011,Papadogiannis:2008,Huang:2011}. The performance of multicell cooperation schemes designed under ideal conditions (ideal CSI measurement and feedback, ideal link adaptation, no delay, unlimited backhaul, infinite overhead, no dynamic interference, full knowledge of receiver implementation) degrades severely once simulated under more realistic conditions, as evidenced in this paper and in \cite{3GPPTR36.819,Gorokhov:2010}. In particular, the proposed rank coordination is shown to outperform, with a much smaller feedback overhead (only two extra feedback bits) and lower scheduler complexity, the iterative coordinated scheduling and beamforming of \cite{Yu:2011} in a realistic setup with non-ideal feedback and link adaptation.

\item The reported information accounts for the receiver interference suppression capability and the effect of cooperation while deriving the CQI, the serving and the interference rank. This helps the BS to select the appropriate modulation and coding level and benefit from link adaptation. Moreover the coordinated scheduler can be designed and operated accounting for the fact that the reported information accounts for the contribution of the receiver in mitigating the ICI. This contrasts with the aforementioned schemes relying on explicit feedback, e.g.\ \cite{Dahrouj:2010,Cadambe:2008,Choi:2008,Kiani:2008,Gjendemsjoe:2008,Huang:2006,Venturino:2010,Venturino:2009,Yu:2010,Yu:2011,Papadogiannis:2008,Tajer:2011,Lee:2011,Huang:2011}, where the BSs have to compute the CQI based on the (full) CSI feedback. To do so, it is assumed that the BSs involved in the iterative scheduler know the characteristics of user terminals (e.g.\ receiver ability to cancel inter-cell interference). However such characteristics are specific to the terminal implementation and are not shared with the BSs in any practical system, therefore making the link adaptation challenging with the iterative scheduler. As discussed in the evaluation section of this paper, this issue is especially true in the presence of non-ideal feedback where the computed transmission rank and CQIs at the BS easily mismatch with the actual supportable transmission rank and SINR.  

\item The user report is applicable to closed-loop and open-loop MIMO operations, i.e.\ irrespectively of whether the Precoder Matrix Indicator (PMI) is reported or not \cite{Li:2010}.
\end{itemize}

\par \textit{At the network side}, with the use of an appropriate coordinated scheduler motivated by an interference pricing mechanism similar to \cite{Huang:2006} and relying on a Master-Slave architecture, cells coordinate with each other to take informed decisions on the scheduled users and the transmission ranks that would be the least detrimental to the victim users in the neighboring cells. 
In the iterative scheduler, multiple iterations are required to converge (if convergence is achieved). The final scheduling decisions are obtained only after a very long latency as every iteration requires to wait for the user report. For reference, \cite{Yu:2010} requires approximately 500 iterations (50 iterations where each iteration consists of 10 sub-iterations) before convergence to coordinate power among cells. Those extensive interactions between the users and the BSs significantly increase the complexity and the overhead of the network as well as the synchronization and backhaul requirements, making it not practical. 
The Master-Slave coordinated scheduler on the other hand operates in a more distributed manner and relies only on some low-overhead inter-cell message exchange. It is moreover less sensitive to convergence problems. 

\par The last few paragraphs highlight a fundamental difference in system design between the referred coordination schemes (relying on explicit feedback) and the proposed rank coordination: while the former puts all the coordination burden on the network side, the later decreases the coordination burden at the network side by bringing the contribution of the receivers into the multi-cell coordination. Thereby, the rank coordination scheme balances the overall effort of multi-cell coordination between the receivers and the network. To do so, the receivers are not supposed to simply report CSI but act smartly by making appropriate recommendation (in the form of a report of a preferred interference rank computed accounting for the receiver interference rejection capability) to the network. While the later approach may not be helpful in ideal situations because the network possesses all necessary information to make accurate decisions, it becomes particularly helpful when the aim is to design multi-cell coordination schemes for non-ideal setup (when the network does not have enough information to make accurate decisions).

\par The paper is organized as follows. Section \ref{system_model} details the system model and section \ref{resource_allocation} formulates the resource allocation problem and derives the guidelines for the coordinated scheduler design. The principles and implementation details of the rank recommendation-based coordinated scheduling are described in sections \ref{RR_coordinated_scheduling} and \ref{practical_implementation}, respectively. Section \ref{SLS} illustrates the achievable gains of the proposed scheme based on system level evaluations.

\section{System Model}\label{system_model}

We assume a downlink multi-cell MIMO-OFDMA network with a total number of $K$ users distributed in $n_c$ cells, with $K_i$ users in every cell $i$, $T$ subcarriers, $N_t$ transmit antennas at every BS, $N_r$ receive antenna at every mobile terminal.

\par Assume that the MIMO channel between cell $i$ and user $q$ on subcarrier $k$ writes as $\alpha_{q,i}^{1/2} \mathbf{H}_{k,q,i}$ where $\mathbf{H}_{k,q,i} \in \mathbb{C}^{N_r \times N_t}$ models the small scale fading process of the MIMO channel and $\alpha_{q,i}$ refers to the large-scale fading (path loss and shadowing). Note that the large-scale fading is typically independent of the subcarrier.

\par The \textit{serving cell} is defined as the cell transmitting the downlink control information. We define the \textit{served user set} of cell $i$, denoted as $\mathcal{K}_i$ with cardinality $\sharp\mathcal{K}_i=K_i$, as the set of users who have cell $i$ as serving cell. We also define the \textit{scheduled user set} of cell $i$ on subcarrier $k$, denoted as $\mathbf{K}_{k,i}\subset\mathcal{K}_i$, as the subset of users $\in\mathcal{K}_i$ who are actually scheduled on subcarrier $k$ at a certain time instant.

\par In this paper, for the sake of readability, we assume single-user transmissions (i.e.\ a single user is allocated on a given time and frequency resource). Therefore, the cardinality of $\mathbf{K}_{k,j}$ $\forall j$ is always equal to 1. On subcarrier $k$, cell $i$ serves the user belonging to $\mathbf{K}_{k,i}$ with $L_{k,i}$ data streams ($1 \leq L_{k,i}\leq N_t$). The transmit symbol vector $\mathbf{x}_{k,i} \in \mathbb{C}^{L_{k,i}}$ made of $L_{k,i}$ symbols is power controlled by $\mathbf{S}_{k,i}\in \mathbb{R}^{L_{k,i} \times L_{k,i}}$ and precoded by the transmit precoder $\mathbf{F}_{k,i} \in \mathbb{C}^{N_t \times L_{k,i}}$ such that the transmit precoded symbol vector writes as $\tilde{\mathbf{x}}_{k,i}=\mathbf{F}_{k,i} \mathbf{S}_{k,i}^{1/2} \mathbf{x}_{k,i}$. $\mathbf{F}_{k,i}$ is made of $L_{k,i}$ columns denoted as $\mathbf{f}_{k,i,m}$, $m=1,...,L_{k,i}$. $\mathbf{F}_{k,i}$ can refer to either a closed-loop precoder designed based on the CSI feedback or an open-loop precoder pre-defined per transmission rank $L_{k,i}$ (e.g.\ space-time/frequency code or open-loop Single-User spatial multiplexing). Note that, while we assume SU-MIMO transmission for the sake of readability, the rank coordination can be extended to a multi-user MIMO set-up.

\par For the user $q\in\mathbf{K}_{k,i}$ scheduled in cell $i$ on subcarrier $k$, the received signal $\tilde{\mathbf{y}}_{k,q} \in \mathbb{C}^{N_r}$ is shaped by $\mathbf{G}_{k,q}\in \mathbb{C}^{L_{k,i} \times N_r}$ and the filtered received signal $\mathbf{y}_{k,q} \in \mathbb{C}^{L_{k,i}}$ writes as 
\begin{align}
\mathbf{y}_{k,q}=\mathbf{G}_{k,q} \tilde{\mathbf{y}}_{k,q}=\sum_{j=1}^{n_c}\alpha_{q,j}^{1/2} \mathbf{G}_{k,q}\mathbf{H}_{k,q,j}\mathbf{F}_{k,j}\mathbf{S}_{k,j}^{1/2}\mathbf{x}_{k,j}+\mathbf{n}_{k,q}
\end{align}
where $\mathbf{n}_{k,q}=\mathbf{G}_{k,q}\tilde{\mathbf{n}}_{k,q}$ and $\tilde{\mathbf{n}}_{k,q}$ a complex Gaussian noise $\mathbb{CN}\big(0,\sigma_{n,k,q}^2\mathbf{I}_{N_r}\big)$. The receive filter $\mathbf{G}_{k,q}$ is made of $L_{k,i}$ rows denoted as $\mathbf{g}_{k,q,m}$, $m=1,...,L_{k,i}$. The strategy to compute $\mathbf{G}_{k,q}$ is assumed to be only known by the receiver and not by the transmitter (similarly to practical systems). Examples of strategies include MMSE with ideal or simplified ICI rejection capabilities (as used in the evaluations in Section \ref{SLS}). In this paper, similarly to practical systems as LTE-A, we will assume uniform power allocation among streams, i.e. $\mathbf{S}_{k,i}=E_{s,i}/L_{k,i}$ where $E_{s,i}$ is the total transmit power at BS $i$. 

\par The variable $\mathbf{K}$ collects the user assignments for all subcarriers and all cells and writes as $\mathbf{K}=\left\{\mathbf{K}_{i}\right\}_{i=1}^{n_c}$ where $\mathbf{K}_{i}=\left\{\mathbf{K}_{k,i}\right\}_{\forall k}$. Similarly, we define  $\mathbf{L}=\left\{\mathbf{L}_{i}\right\}_{i=1}^{n_c}$ where $\mathbf{L}_{i}=\left\{L_{k,i}\right\}_{\forall k}$.

\par In order to ease explanations, we define the CoMP measurement set in analogy with 3GPP terminology \cite{3GPPTR36.819}. The \textit{CoMP measurement set} of user $q\in\mathcal{K}_i$, whose serving cell is $i$, is defined as the set of cells about which channel state/statistical information related to their link to the user is reported to the BS and is expressed based on long term channel properties as
\begin{equation}\label{measurement_set}
\mathcal{M}_q=\left\{\Big.j\Big|\frac{\alpha_{q,i}}{\alpha_{q,j}}<\delta,\forall j\neq i\right\}
\end{equation}
for some threshold $\delta$. The larger $\delta$, the larger the CoMP measurement set and the higher the feedback overhead.
As defined, the CoMP measurement set does not include the serving cell $i$. Hence to operate multi-cell cooperation, a user feeds back its serving cell CSI and the CoMP measurement set CSI. We denote by a CoMP user, a user whose CoMP measurement set is not empty. The \textit{CoMP users set} of cell $i$ is defined as $\mathcal{P}_i=\left\{q\in\mathcal{K}_i\left.\right|\mathcal{M}_q\neq \emptyset\right\}$.

\par The \textit{CoMP-requested user set} of cell $i$ is defined as the set of users that have cell $i$ in their CoMP measurement set, i.e.\
$\mathcal{R}_i=\left\{\left.l\right|i\in\mathcal{M}_l, \forall l\right\}$.
Note that the CoMP-requested user set can also be viewed as the victim user set of cell $i$ as it is the set of users who could be impacted by cell $i$ interference.

\section{Coordinated multi-cell resource allocation}\label{resource_allocation}

\par Contrary to a non-cooperative network, a cooperative scheme relying on rank coordination coordinates dynamically the users in all cells and frequency resources such that the transmission rank of a given cell and frequency resource is favorable to the performance of that cell's users and of the adjacent cells' victim users scheduled on the same frequency resource. In this section, the resource allocation problem related to rank coordination is discussed and some scheduler architecture motivated by an interference pricing mechanism is introduced.

\par We make the following assumption in this section. 
\begin{assumption}\label{assumption_pricing}
The transmission rank $L_{k,j}$ ${\forall j}$ is a real variable and the throughput $T_{k,q,i}$ of user $q$ in cell $i$ on subcarrier $k$ is a continuous function of $\left\{L_{k,j}\right\}_{\forall j}$. The beamforming directions $\mathbf{F}_{k,j}$ are fixed and predefined for every transmission rank $L_{k,j}$, $\forall j$.
\end{assumption}
As it will apear clearer in the sequel, this assumption is used to relax the optimization problem (by dealing with real rather than integer transmission ranks). Under assumption \ref{assumption_pricing}, we motivate the guidelines of the scheduler architecture of section \ref{RR_coordinated_scheduling}. The practical implementation of the scheduler dealing with integer transmission ranks and variable beamforming directions is addressed in Section \ref{practical_implementation}.

\subsection{Problem Statement}\label{problem_statement}

\par We denote and define the weighted rate of cell $i$ on subcarrier $k$ as $T_{k,i}=w_{q}T_{k,q,i}$ where $q\in\mathbf{K}_{k,i}$. The weights $w_{q}$ account for fairness among users (and may be related for instance to the QoS of each user) and $T_{k,q,i}$ refers to the rate of scheduled user $q$ in cell $i$ on subcarrier $k$. At this stage, we view $T_{k,q,i}$ and $T_{k,i}$ as abstract functions of the transmission rank in each cell. Hence we sometimes denote explicitly $T_{k,q,i}\big(\left\{L_{k,j}\right\}_{\forall j}\big)$ and $T_{k,i}\big(\left\{L_{k,j}\right\}_{\forall j}\big)$.
\par The problem is to maximize the network weighted sum-rate accounting for fairness among users and cells and design a coordinated scheduler that decides which frequency resource to allocate to which user in every cell with the appropriate transmission rank. We write
\begin{equation}\label{scheduling_problem_MIMO_OFDMA}
\left\{\mathbf{K}^{\star},\mathbf{L}^{\star}\right\}=\arg\max_{\mathbf{K}\subset\mathcal{K},\mathbf{L}} \sum_{j=1}^{n_c}\sum_{k=0,q\in\mathbf{K}_{k,j}}^{T-1}w_{q}T_{k,q,j}.
\end{equation}
Given the uniform power allocation and the assumption \ref{assumption_pricing} on the fixed beamformers, the problem \eqref{scheduling_problem_MIMO_OFDMA} is to be maximized over transmission ranks and user schedule only.

\par At a first glance, problem \eqref{scheduling_problem_MIMO_OFDMA} could be viewed as a sub-problem of the more general problem of a joint coordinated scheduling, beamforming and power control \cite{Yu:2011}. As explained in the introduction, we resort to an alternative way of solving \eqref{scheduling_problem_MIMO_OFDMA} in order to make the multi-cell cooperation practical. Given that the maximization is performed over the transmission ranks (being integer in a realistic setup) and the user schedule, \eqref{scheduling_problem_MIMO_OFDMA} is a combinatorial problem. Unfortunately, solving such problem would require a centralized architecture that is not desirable \cite{Kiani:2008,Gjendemsjoe:2008,Kiani:2009}. By relaxing the transmission ranks being integer to real, we can motivate the use of a distributed and practical scheduler architecture. Following assumption \ref{assumption_pricing}, we therefore assume in the maximization problem \eqref{scheduling_problem_MIMO_OFDMA} that the transmission ranks $\mathbf{L}$ are real and subject to the constraints $L_{k,j}\geq L_{min,j}$ and $L_{k,j}\leq L_{max,j}$. $L_{min,j}$ and $L_{max,j}$ refer to the minimum and maximum transmission rank in cell $j$, respectively and could be configured by the network (typically, $L_{min,j}=1$ and $L_{max,j}=N_t$).
\par The proposed architecture relies on a Master-Slave distributed architecture and interference rank recommendation motivated by the derivations of the next section. Performance evaluations in Section \ref{SLS} will demonstrate the benefits of the rank recommendation compared to the heavy machinery of the iterative coordinated scheduling, beamforming and power control in a realistic setup.

\subsection{Motivations for the scheduler architecture}\label{interference_pricing}

\par For a fixed user schedule, the optimal rank allocation problem must satisfy the Karush-Kuhn-Tucker (KKT) conditions. The Lagrangian of the optimization problem dualized with respect to the rank constraint writes as
\begin{multline}
\mathcal{L}\left(\mathbf{K},\mathbf{L},\mathbf{\nu},\mathbf{\mu}\right)=\sum_{j=1}^{n_c}\sum_{k=0}^{T-1}\left[T_{k,j}
+\nu_{k,j}\left(L_{max,j}-L_{k,j}\right)\right.\\\left.+\mu_{k,j}\left(L_{k,j}-L_{min,j}\right)\right]
\end{multline}
where $\mathbf{\nu}=\left\{\nu_{k,j}\right\}_{k,j}$ and $\mathbf{\mu}=\left\{\mu_{k,j}\right\}_{k,j}$ are the sets of non-negative Lagrange multipliers associated with the transmission rank constraints in each cell and each subcarrier.

For any $i=1,\ldots,n_c$ and $k=0,\ldots,T-1$, the solution should satisfy
\begin{equation}\label{derive_Lagrangian_MIMO}
\frac{\partial\mathcal{L}}{\partial L_{k,i}}=0,
\end{equation}
$\nu_{k,i}\left(L_{max,i}-L_{k,i}\right)=0$, $\mu_{k,i}\left(L_{k,i}-L_{min,i}\right)=0$, $\nu_{k,i}\geq 0$ and $\mu_{k,i}\geq 0$.

We can proceed with \eqref{derive_Lagrangian_MIMO} as
\begin{equation}\label{derive_Lagrangian_pricing_MIMO}
\frac{\partial T_{k,i}}{\partial L_{k,i}}-\sum_{m\neq i, s\in\mathbf{K}_{k,m}}w_{s}\pi_{k,s,m,i}=\nu_{k,i}-\mu_{k,i}
\end{equation}
where we define 
\begin{equation}\label{Pi_kqi}
\pi_{k,s,m,i}=-\frac{\partial T_{k,s,m}}{\partial L_{k,i}}.
\end{equation}

\par Let us first define $I_{k,s,i}^{\star}$ as the transmission rank in cell $i$ that maximizes the throughput $T_{k,s,m}$ of user $s$ in cell $m$ assuming a predefined set of transmission ranks $L_{k,j}$ in all cells $j \neq i$
\begin{equation}\label{optimal_rank}
I_{k,s,i}^{\star}=\arg \max_{L_{min,i}\leq L_{k,i}\leq L_{max,i}} T_{k,s,m}\left(L_{k,i},\left\{L_{k,j}\right\}_{j \neq i}\right).
\end{equation}
Note that if the network decides to configure $L_{min,i}=0$, all users will choose their preferred interference rank as being equal to 0, so as not to experience any interference. 

Interestingly, the condition \eqref{derive_Lagrangian_pricing_MIMO} can be viewed as the KKT condition of the problem where each cell $i$ tries to maximize on subcarrier $k$ the following surplus function
\begin{equation}\label{surplus_function_MIMO}
\Upsilon_{k,i}=T_{k,i}-\Pi_{k,i}
\end{equation}
with
\begin{equation}\label{PI_payment}
\Pi_{k,i}=\sum_{m\neq i}\sum_{s\in\mathbf{K}_{k,m}}\left(L_{k,i}-I_{k,s,i}^{\star}\right)w_{s}\pi_{k,s,m,i},
\end{equation}
assuming fixed $L_{k,j}$ with $j \neq i$, $I_{k,s,i}^{\star}$ and $\pi_{k,s,m,i}$ with $\left(s,m\right)\neq \left(q,i\right)$.

\par Equation \eqref{surplus_function_MIMO} has an interference pricing interpretation, with some similarities with the interference pricing mechanism introduced for power control in \cite{Huang:2006,Yu:2010}. Here, we show that a similar pricing mechanism can be used to proceed with another type of coordination, namely rank coordination rather than power control. Indeed, given \eqref{optimal_rank}, we can safely write that, in the vicinity of $I_{k,s,i}^{\star}$, the throughput $T_{k,s,m}$ of user $s$ in cell $m$ writes as a concave function of $L_{k,i}$, i.e.\
$\frac{\partial T_{k,s,m}}{\partial L_{k,i}} \geq  0 $ if $L_{k,i}\leq I_{k,s,i}^{\star}$ and $\frac{\partial T_{k,s,m}}{\partial L_{k,i}} \leq  0$ if $L_{k,i}\geq I_{k,s,i}^{\star}$.
Under such an assumption, $\big(L_{k,i}-I_{k,s,i}^{\star}\big)\pi_{k,s,m,i}$ and $\Pi_{k,i}$ are non-negative. $\Upsilon_{k,i}$ is the weighted sum-rate in cell $i$ minus the payment $\Pi_{k,i}$ due to the interference created to the victim users scheduled in the neighboring cells. 
\par The payment $\Pi_{k,i}$ accounts for the weighted sum of all prices $\pi_{k,s,m,i}$ over all scheduled users $s$ in the network. The weight of a given user is proportional to its QoS and the deviation of the actual transmission rank in cell $i$ with respect to the transmission rank in cell $i$ that would maximize the victim user $s$ throughput in cell $m$. If such a deviation is null for a certain user $s$, cell $i$ is not fined for the interference created to user $s$. The price $\pi_{k,s,m,i}$ refers to how much the throughput of user $s$ in cell $m$ is sensitive to any change of the transmission rank of cell $i$. The quantity $\tilde{w}_{k,s,i}=w_s\pi_{k,s,m,i}$ can be thought of as the overall sensitivity of user $s$ to any deviation of the transmission rank in cell $i$ from its optimal $I_{k,s,i}^{\star}$ and we can equivalently write the payment as $\Pi_{k,i}=\sum_{m\neq i,s\in\mathbf{K}_{k,m}}\big(L_{k,i}-I_{k,s,i}^{\star}\big)\tilde{w}_{k,s,i}$.

\par Equation \eqref{surplus_function_MIMO} suggests that the cell $i$ can decide upon the set of co-scheduled users and the transmission rank on subcarrier $k$ as follows
\begin{align}\label{user_selection_iterative_algo_MIMO}
\left\{\mathbf{K}_{k,i}^{\star},L_{k,i}^{\star}\right\}=\arg\max_{\mathbf{K}_{k,i},L_{k,i}}\Upsilon_{k,i}.
\end{align}

\section{Rank recommendation-based coordinated scheduling}\label{RR_coordinated_scheduling}

\par Motivated by the interference pricing mechanism, we derive in this section some guidelines for the rank recommendation-based coordinated scheduler that coordinates transmission ranks and scheduled users in the network and compute the locally (hopefully) optimum $\mathbf{L}^{\star}$ and $\mathbf{K}^{\star}$ based on the recommendations made by the terminals. From \eqref{surplus_function_MIMO}, we make the following first observation.

\begin{observation}
The coordinated scheduler in cell $i$ has to rely on the report of some local CSI from terminals $\in\mathcal{K}_i$ to perform single-cell processing at the BS and compute the term $T_{k,i}=w_{q} T_{k,q,i}$, $q\in\mathbf{K}_{k,i}$. It also relies on some message exchanges between cells, namely the reception by cell $i$ of the price information $w_{s}\pi_{k,s,m,i}$ and $I_{k,s,i}^{\star}$ for all $s\in\mathcal{R}_i$ and the transfer from cell $i$ of the price information $w_{q}\pi_{k,q,i,j}$ and $I_{k,q,j}^{\star}$ for all $q\in\mathcal{P}_i$ and $j\in\mathcal{M}_q$. 
\end{observation}

\par In a classical explicit feedback approach (as used in the multi-cell coordination techniques of \cite{Yu:2010,Yu:2011}), quantities like $T_{k,q,i}$, $I_{k,s,i}^{\star}$, $I_{k,q,j}^{\star}$, $\pi_{k,s,m,i}$ and $\pi_{k,q,i,j}$ would be computed at the BS based on the CSI feedback and assuming that the receiver implementation is known to the BS. However, as explained in the introduction, the accurate computations of those quantities 
are very challenging at the BS side as they are a function of many parameters specific to the receiver implementation and are highly sensitive to the accuracy of the channel measurement and feedback. In order to bring the contribution of the receiver in the design of the coordinated scheduler, it is preferable that the user terminal $q$ (and similarly for terminal $s$) estimates, computes and reports $T_{k,q,i}$, $I_{k,q,j}^{\star}$ and $\pi_{k,q,i,j}$ by accounting for the transmission ranks in the interfering cells, its receiver interference rejection capability and the measured channels as perceived at the receiver sides.

\par Focusing on cell $i$, the terminals $q\in\mathbf{K}_{k,i}$ and $s\in\mathcal{R}_i$ and cell $i$ scheduler cooperate with the aim of maximizing $\Upsilon_{k,i}$ in \eqref{surplus_function_MIMO} and decreasing $\Pi_{k,i}$ in \eqref{PI_payment}.

\begin{observation}
In order to help cell $i$ scheduler, any user $q\in\mathcal{K}_{i}$ served by cell $i$ reports an estimate of $T_{k,q,i}$ and any user $s\in\mathcal{R}_i$ belonging to a cell $m$, victim of cell $i$ interference, recommends cell $i$ to choose $L_{k,i}=I_{k,s,i}^{\star}$. User $s$ reports targeting cell $i$ contain $I_{k,s,i}^{\star}$ and an estimate of the user throughput loss $\Delta T_{k,s,i}$ achievable if the recommendation is not accounted for in cell $i$ decisions on the transmission ranks. 
\end{observation}

The report of the user throughput loss, defined as $\Delta T_{k,s,i}=T_{k,s,m}\big(L_{k,i},\left\{L_{k,j}\right\}_{j \neq i}\big)-T_{k,s,m}\big(I_{k,s,i}^{\star},\left\{L_{k,j}\right\}_{j \neq i}\big)$ for some predefined $\left\{L_{k,j}\right\}_{j \neq i}$, enables cell $i$ to compute the price as follows  
$\pi_{k,s,m,i}\approx -\frac{\Delta T_{k,s,i}}{L_{k,i}-I_{k,s,i}^{\star}}$. The quantity $\big(L_{k,i}-I_{k,s,i}^{\star}\big)\pi_{k,s,m,i}$ expresses the variation in user $s$ throughput due to the transmission rank $L_{k,i}$ rather than $I_{k,s,i}^{\star}$. 

\par On the network side, the scheduler in cell $i$ strives to respect as much as possible the recommendation of the CoMP users and guarantee $L_{k,i}-I_{k,s,i}^{\star}=0$ on subcarriers where the victim user $s\in\mathbf{K}_{k,m}$ of cell $i$ is scheduled, as highlighted by the following observation.
\begin{observation}\label{lesson_scheduler}
Whenever the scheduler of a given cell $i$ accepts the request of a recommended interference rank $I_{k,s,i}^{\star}$ at time instant $t$ and over frequency resource $k$, the victim user $s$ in the neighboring cell $m$ who reported the recommended interference rank $I_{k,s,i}^{\star}$ to cell $i$ has to be scheduled at the same time instant $t$ and on the same frequency resource $k$.
\end{observation} 

\section{Practical implementation}\label{practical_implementation}

In this section, we exploit the observations made in previous section and come up with some practical implementation of the rank recommendation-based coordinated scheduling. In particular, we drop the assumption \ref{assumption_pricing} and discuss the effect of variable beam directions.

\subsection{Wideband rank recommendation}

\par Practical systems rely on rank indicator (RI), CQI and Precoding Matrix Indicator (PMI) reports \cite{Li:2010}. RI commonly refers to the preferred serving cell transmission rank and is a wideband and potentially long term information as it changes relatively slowly in the frequency and time domains. RI report therefore incurs a very small feedback overhead. As for now, any reported rank information in the proposed scheme will be wideband, while CQI and PMI are subband information.
\par For a CoMP user $q$ associated with the serving cell $i$ ($q\in\mathcal{K}_i$) and victim of a cell $j\in\mathcal{M}_q$, this terminal reports its preferred serving cell wideband RI $R_{q}^{\star}$, i.e.\ the user makes the hypothesis that $L_{k,i}=R_{q}^{\star}$ $\forall k$ at the time of report and that $R_{q}^{\star}$ maximizes user $q$ throughput \cite{Li:2010}. The same user $q$ also transmits to the serving cell $i$ the transmission rank of the interfering cell $j\in\mathcal{M}_q$, denoted as the preferred interference RI $I_{q,j}^{\star}$, that maximizes its performance. The user recommends the interfering cell $j$ to transmit a number of streams corresponding to\footnote{Note that we refer to $I_{q,j}^{\star}$ rather than $I_{k,q,j}^{\star}$ as in previous sections to stress the fact that the preferred interference RI is a wideband information.} $I_{q,j}^{\star}$, i.e.\ $L_{k,j}=I_{q,j}^{\star}$ $\forall k$. 

\subsection{Computation of the preferred interference rank}\label{interf_rank_computation}

In Section \ref{resource_allocation}, fixed beamforming directions and real transmission ranks are assumed. However, the user $q\in\mathcal{K}_i$ does not know the precoder in the interfering cell $j$ at the time of CQI, $R_{q}^{\star}$ and $I_{q,j\in\mathcal{M}_q}^{\star}$ reports. In order to cope with such issue, similarly to the channel information partitioning strategy in \cite{Kiani:2009}, the terminal computes the required information by averaging the throughput over the possible realizations of the transmit precoder $\mathbf{F}_{k,j}$ in the interfering cells $j\in\mathcal{M}_q$, given the current realization of the channel matrices (measured at the terminal). Those precoders can be assumed to be selected in the limited feedback codebook $\mathcal{C}$ (defined for each rank and assumed the same in all cells) and the throughput average can be computed for each set of serving cell rank $L_{k,i}$, precoder $\mathbf{F}_{k,i}$ and interference rank $\left\{L_{k,j}\right\}_{j\in\mathcal{M}_q}$
\begin{equation}
\tilde{T}_{k,q,i}\left(\mathbf{F}_{k,i},L_{k,i},\left\{L_{k,j}\right\}_{j\in\mathcal{M}_q}\right)\approx\mathcal{E}_{\left\{\mathbf{F}_{k,j}\in\mathcal{C}\right\}_{j\in\mathcal{M}_q}}\left\{T_{k,q,i}\right\}
\end{equation}
where 
\begin{equation}
T_{k,q,i}= \sum_{m=1}^{L_{k,i}}\log_2\left(1+\rho_{k,q,m}\right)
\end{equation}
with
\begin{equation}
\rho_{k,q,m}=\frac{\alpha_{q,i}\left|\mathbf{g}_{k,q,m}\mathbf{H}_{k,q,i}\mathbf{f}_{k,i,m}\right|^2 E_{s,i}/L_{k,i}}{\sum_{j\in\mathcal{M}_q}\alpha_{q,j}\left\|\mathbf{g}_{k,q,m}\mathbf{H}_{k,q,j}\mathbf{F}_{k,j}\right\|^2 E_{s,i}/L_{k,j}+\sigma_{n,k,q}^2}.
\end{equation}
The computation of $\tilde{T}_{k,q,i}$ accounts for the receive filter $\mathbf{G}_{k,q}$ and therefore the interference rejection capability of the receiver.

\par Following \eqref{optimal_rank}, the user $q$ in cell $i$ can jointly compute the best set of preferred serving cell RI $R_{q}^{\star}$ and preferred recommended interference RI $I_{q,j}^{\star}$, as follows
\begin{equation}\label{selection_ranks_average_rate}
\left\{R_{q}^{\star},\left\{I_{q,j}^{\star}\right\}_{j\in\mathcal{M}_q}\right\}=\arg\max_{L_{k,i},\left\{L_{k,j}\right\}_{j\in\mathcal{M}_q}}\;\mathcal{E}_{k}\left\{\max_{\mathbf{F}_{k,i}\in\mathcal{C}}\tilde{T}_{k,q,i}\right\},
\end{equation}
where the averaging is performed over all subcarriers due to the wideband report of the RIs and the maximization is done over a restricted set of integers $L_{k,j}\in\left\{L_{min,j},\ldots,L_{max,j}\right\}$ $\forall j$. For a given set of transmission ranks $L'_{k,i},\left\{L'_{k,j}\right\}_{j\in\mathcal{M}_q}$, the best precoder (for closed-loop operations) for user $q$ in cell $i$ on subcarrier $k$ is selected as
\begin{multline}
\mathbf{F}_{k,i}^{\star}\left(L'_{k,i},\left\{L'_{k,j}\right\}_{j\in\mathcal{M}_q}\right)\\=\arg \max_{\mathbf{F}_{k,i}\in\mathcal{C}}\tilde{T}_{k,q,i}\left(\mathbf{F}_{k,i},L'_{k,i},\left\{L'_{k,j}\right\}_{j\in\mathcal{M}_q}\right).
\end{multline}

\par Once $R_{q}^{\star}$ and $I_{q,j\in\mathcal{M}_q}^{\star}$ are selected, the estimate of user $q$ throughput to be reported to the network is given by $\tilde{T}_{k,q,i}^{\star}=\tilde{T}_{k,q,i}\big(\mathbf{F}_{k,i}^{\star}\big(R_{q}^{\star},\left\{I_{q,j}^{\star}\right\}_{j\in\mathcal{M}_q}\big),R_{q}^{\star},\left\{I_{q,j}^{\star}\right\}_{j\in\mathcal{M}_q}\big)$ while the estimate of the throughput loss writes as $\Delta\tilde{T}_{k,q,i}=\tilde{T}_{k,q,i}\big(\mathbf{F}_{k,i}^{\star}\big(L_{k,i},\left\{L_{k,j}\right\}_{j\in\mathcal{M}_q}\big),L_{k,i},\left\{L_{k,j}\right\}_{j \neq i}\big)-\tilde{T}_{k,q,i}^{\star}$, $\forall \left\{L_{k,i},\big\{L_{k,j}\big\}_{j \neq i}\right\}\neq\left\{R_{q}^{\star},\big\{I_{q,j}^{\star}\big\}_{j\in\mathcal{M}_q}\right\}$.
Given the user reports of $R_{q}^{\star}$, $\left\{I_{q,j}^{\star}\right\}_{j\in\mathcal{M}_q}$, $\tilde{T}_{k,q,i}^{\star}$, $\big\{\Delta\tilde{T}_{k,q,i}\big\}$ and $\mathbf{F}_{k,i}^{\star}\big(R_{q}^{\star},\left\{I_{q,j}^{\star}\right\}_{j\in\mathcal{M}_q}\big)$ (for closed-loop operations), the coordinated scheduler can estimate the surplus function \eqref{surplus_function_MIMO} with the objective of performing \eqref{user_selection_iterative_algo_MIMO}. In a practical system,  $\tilde{T}_{k,q,i}^{\star}$ and $\Delta\tilde{T}_{k,q,i}$ would be reported using a CQI and a differential (also called delta) CQI, respectively. We will without loss of generality and for simplicity denote them as CQI and differential CQI in the sequel.

\par Note that the selection of the preferred interference rank highly depends on the receiver architecture. While an inter-cell interference rejection combiner would favor lower interference rank, it is not so necessarily the same for other types of receivers.

\subsection{A Master-Slave scheduler architecture}\label{Master_Slave}

\par The coordinated scheduler relies on an asynchronous Master-Slave architecture motivated by Observation \ref{lesson_scheduler}. At each time instant, only one BS acts as the Master (denoted as M) and the other BSs are the slave (denoted as S). The Master BS, based on the reports of the preferred interference rank, decides a certain transmission rank $L_{k,\textrm{M}}$ constant $\forall k$, i.e.\ $L_{k,\textrm{M}}=L_{\textrm{M}}$, and schedules its users such that the transmission ranks of all scheduled users are as much as possible equal to $L_{\textrm{M}}$. The Slave BSs, knowing that the Master BS will accept some recommended interference rank, will schedule with highest priority their CoMP users who requested rank coordination to the Master BS. 

\par Assume for ease of presentation and without loss of generality that a cluster is made of 3 cells (e.g.\ as in intra-site deployments) \cite{Gesbert:2010,Papadogiannis:2008}. Table \ref{Master_slave} illustrates the operation of the scheduler for such a 3-cells cluster. For a given time instant, there are one Master BS (denoted as M) and two slave BSs (denoted as S$_{\textrm{1}}$ and S$_{\textrm{2}}$).

% Table 1
\begin{table*}
\caption{Example of the Master-Slave scheduler architecture}
\centering
\begin{tabular}{|c||c|c|c|c|c|c|c|c|c|}
\hline time & 1 & 2 & 3 & 4 & 5 & 6 & 7 & 8 & 9  \\
\hline
\hline BS$_{\textrm{1}}$ & M, $L_M$=2  & S$_{\textrm{1}}$ & S$_{\textrm{1}}$ & M, $L_M$=1  & S$_{\textrm{1}}$ & S$_{\textrm{1}}$ & M, $L_M$=2  & S$_{\textrm{1}}$ & S$_{\textrm{1}}$  \\
\hline BS$_{\textrm{2}}$ & S$_{\textrm{1}}$ & M, $L_M$=1  & S$_{\textrm{2}}$ & S$_{\textrm{1}}$ & M, $L_M$=2 & S$_{\textrm{2}}$ & S$_{\textrm{1}}$ & M, $L_M$=1  & S$_{\textrm{2}}$ \\
\hline BS$_{\textrm{3}}$ & S$_{\textrm{2}}$ & S$_{\textrm{2}}$ & M, $L_M$=3 & S$_{\textrm{2}}$ & S$_{\textrm{2}}$ & M, $L_M$=1 & S$_{\textrm{2}}$ & S$_{\textrm{2}}$ & M, $L_M$=3 \\
\hline
\end{tabular}
\label{Master_slave}
\end{table*}

\subsubsection{Master BS decision on the transmission rank}

\par The Master BS, upon reception of all information\footnote{Using the same notation as in previous section, the interference ranks recommended to interfering cell M by users $l$ in S$_{\textrm{1}}$ and S$_{\textrm{2}}$ are denoted as $I_{l,\textrm{M}}^{\star}$ with $l\in\left\{\mathcal{K}_{\textrm{S}_\textrm{1}},\mathcal{K}_{\textrm{S}_\textrm{2}}\right\}$.} $I_{l,\textrm{M}}^{\star}$ and all the effective QoS $\tilde{w}_{k,l,M}$ of victim users $l$, with $l\in\left\{\mathcal{K}_{\textrm{S}_\textrm{1}},\mathcal{K}_{\textrm{S}_\textrm{2}}\right\}$, sorts those interference ranks by order of priority. 
In a given cell $i$, the vector $I_{1}^{(i)},I_{2}^{(i)},I_{3}^{(i)},...,I_{N}^{(i)}$ denotes the priority of the interference ranks. For instance, [$I_{1}^{(1)}$,$I_{2}^{(1)}$,$I_{3}^{(1)}$,$I_{4}^{(1)}$]=[2,1,3,4] indicates that a recommended interference rank equal to 2 is the most prioritized in cell 1. Master BS M decides upon the transmission rank $L_{\textrm{M}}$ and allocates one transmission rank for each subframe where the BS acts as a Master BS. By doing so the each Master BS defines a cycling pattern of the transmission ranks with the objective of guaranteeing some time-domain fairness. 
The priority and allocation of the transmission ranks accounts for the relative number of rank recommendation requests per rank, for the QoS $w_l$ and the delta CQI (or equivalently the effective QoS $\tilde{w}_{k,l,M}$) of victim CoMP users $l$ in S$_{\textrm{1}}$ and S$_{\textrm{2}}$ and for the QoS of cell M users. In its simplest version used in the evaluation section \ref{SLS}, the priority is exclusively determined based on the relative number of rank recommendation requests per rank. 
\par Let us illustrate the operation through the example on Table \ref{Master_slave}. The value of $L_{\textrm{M}}$ in a given cell $i$ changes as time (subframe) goes by following the cycling pattern $I_{1}^{(i)}$,$I_{2}^{(i)}$,$I_{1}^{(i)}$,$I_{2}^{(i)}$,$I_{3}^{(i)}$, indicating that whenever cell 1 is the Master BS, BS 1 transmits with rank $L_{\textrm{M}}=I_{1}^{(1)}=2$, $L_{\textrm{M}}=I_{2}^{(1)}=1$, $L_{\textrm{M}}=I_{1}^{(1)}=2$, $L_{\textrm{M}}=I_{2}^{(1)}=1$ and finally $L_{\textrm{M}}=I_{3}^{(1)}=3$ in subframe 1,4,7,10,13 respectively (note that only subframes 1 to 9 are displayed in Table \ref{Master_slave}). BS 2 and 3 operate in a similar manner.

\subsubsection{Master BS scheduler operations}

\par In cell M, we divide users into two subgroups:
\begin{enumerate}
\item $\mathcal{U}_{\textrm{M},1}=\big\{q\in \mathcal{K}_\textrm{M}\left.\right|R_{q}^{\star}=L_{M} \big\}$, i.e.\ the set of users in cell M whose preferred rank indicator is equal to the transmission rank $L_{M}$.
\item $\mathcal{U}_{\textrm{M},2}=\mathcal{K}_\textrm{M}\backslash\mathcal{U}_{\textrm{M},1}=\left\{q\in \mathcal{K}_\textrm{M}\left.\right|q\notin \mathcal{U}_{\textrm{M},1}\right\}$, i.e.\ the other users. 
\end{enumerate}
At a given time instant, the scheduling in cell M is based on proportional fairness (PF) in the frequency domain till all frequency resources are occupied:  
\begin{enumerate}
\item if $\mathcal{U}_{\textrm{M},1}\neq\emptyset$, BS M schedules only users belonging to $\mathcal{U}_{\textrm{M},1}$.
\item if $\mathcal{U}_{\textrm{M},1}=\emptyset$, BS M schedules only users belonging to $\mathcal{U}_{\textrm{M},2}$.  
\end{enumerate}

\subsubsection{Slave BS scheduler operations}

\par In cell $\textrm{S}_i$, $i=1,2$, we define three subgroups:
\begin{enumerate}
\item The set of CoMP users $\in\textrm{S}_i$ who recommend cell M and whose preferred interference rank is equal to the transmission rank $L_M$ as
$\mathcal{U}_{\textrm{S}_i,1}=\left\{q\in \mathcal{P}_{\textrm{S}_i}\left.\right|\textrm{M}\in \mathcal{M}_q,I_{q,\textrm{M}}^{\star}=L_{\textrm{M}} \right\}$.

\item The set of all other CoMP users $\in\textrm{S}_i$, i.e.\ who either do not recommend cell M or recommend cell M but whose preferred interference rank is not equal to the transmission rank, is defined as
$\mathcal{U}_{\textrm{S}_i,2}=\left\{q\in \mathcal{P}_{\textrm{S}_i}\left.\right|\textrm{M}\notin \mathcal{M}_q\right\}\cup \left\{q\in \mathcal{P}_{\textrm{S}_i}\left.\right|\textrm{M}\in \mathcal{M}_q,I_{q,\textrm{M}}^{\star}\neq L_{\textrm{M}} \right\}$.

\item The set of non-CoMP users in $\textrm{S}_i$ is defined as $\mathcal{U}_{\textrm{S}_i,3}=\mathcal{K}_{\textrm{S}_i}\backslash\mathcal{P}_{\textrm{S}_i}$.
\end{enumerate}

Scheduling in cell $\textrm{S}_i$ is performed as follows:
\begin{enumerate}
\item If $\mathcal{U}_{\textrm{M},1}\neq\emptyset$, $\textrm{S}_i$ schedules users in the following order of priority: $\mathcal{U}_{\textrm{S}_i,1}$, $\mathcal{U}_{\textrm{S}_i,3}$ and $\mathcal{U}_{\textrm{S}_i,2}$. 
\item If $\mathcal{U}_{\textrm{M},1}=\emptyset$, $\textrm{S}_i$ schedules all users without any priority (i.e.\ only based on PF constraint).
\end{enumerate}

\subsection{Feedback and Message Passing Requirements}

\par
\begin{figure}[!hhh]
\centerline{\includegraphics[height=\columnwidth,angle=-90]{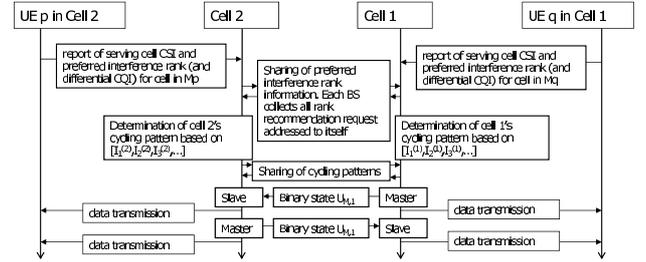}}
  \caption{Overview of the architecture of the rank recommendation based coordinated scheduling.}
  \label{algorithm_overview}
\end{figure}

 Following Observation \ref{lesson_scheduler}, the Master-Slave scheduler guarantees that the transmission rank of cell M, $L_{\textrm{M}}$, equals the preferred recommended interference rank $I_{q,\textrm{M}}^{\star}$ of users $q$ belonging to either $\textrm{S}_1$ or $\textrm{S}_2$ and therefore guarantees that $L_{k,M}-I_{k,q,\textrm{M}}^{\star}=0$ in \eqref{PI_payment} on subcarriers where user $q$ is scheduled. An overview of the architecture of the rank recommendation-based coordinated scheduling is provided in Figure \ref{algorithm_overview}.
We have to note the following important issues. 
\begin{itemize}
\item The serving cell rank, the preferred recommended interference rank, the CQI, PMI and the delta CQI are reported by the users. While the serving cell rank, CQI and PMI stay at the serving cells, the preferred recommended interference rank and the effective QoS (accounting for the delta CQI) are shared among cells. All the rank recommendation requests addressed to a given cell should be collected by that cell. 
We note however that the Master-Slave scheduler mainly relies on the recommended interference rank report. By guaranteeing $L_{k,M}-I_{k,q,\textrm{M}}^{\star}=0$, the tax to be paid by cell M due to the interference created to $\textrm{S}_1$ and $\textrm{S}_2$ decreases significantly (and equals zero in the best case). The report of a delta CQI is mainly useful to adjust (with more fairness) the allocation and the priority of the transmission ranks. It could be skipped to save the feedback overhead.
\item The values of $L_{M}$ need to be shared among cells in the cluster in a periodic manner, i.e.\ $\textrm{S}_1$ and $\textrm{S}_2$ need to be informed about the pattern of transmission ranks e.g.\ $I_{1}^{(i)}$,$I_{2}^{(i)}$,$I_{1}^{(i)}$,$I_{2}^{(i)}$,$I_{3}^{(i)}$ $\forall i$. 
\item $\textrm{S}_1$ and $\textrm{S}_2$ need to be informed dynamically about the binary state $\mathcal{U}_{\textrm{M},1}\neq\emptyset$ or $\mathcal{U}_{\textrm{M},1}=\emptyset$.
\end{itemize}

\par Thanks to the user recommendation, the Master-Slave scheduler architecture does not experience the convergence and complexity issues of the iterative scheduler \cite{Yu:2010,Yu:2011}. It benefits from link adaptation thanks to the computation of a CQI at the user terminal that accounts for multi-cell cooperation and receiver implementation and incurs a very small feedback overhead. Moreover, thanks to the report of the recommended interference rank, a cell edge user $q$ scheduled on resource $k$ can experience higher transmission rank. The appropriate selection of the preferred interference rank $I_{q,j}^{\star}$ enables the user to increase its preferred serving cell rank indicator $R_{q}^{\star}$. Moreover, the wideband RI report is in general robust to the feedback and backhaul delays and to channel estimation errors. 

\section{Performance evaluations}\label{SLS}

\par We compare the performance of closed-loop SU-MIMO with rank adaptation without multi-cell coordination (denoted as SU) and the Master-Slave coordinated scheduler based on rank recommendation (denoted as RR SU). The simulation assumptions (aligned with 3GPP LTE-A \cite{3GPPTR36.819}) are listed in Table \ref{Table_sim_assumptions}. We assume a single wideband preferred serving cell rank indicator and a single wideband recommended interference rank indicator reported every 5ms. The same value of the recommended interference rank for all cells in the CoMP measurement set is used in order to reduce the feedback overhead and simplify the scheduler. This implies that the rank coordination only requires an additional 2 bit feedback overhead (to report the recommended interference rank) compared to the baseline system without coordination (SU). The CQI is computed assuming SU-MIMO transmission as in 3GPP LTE-A for the baseline system and is based on the joint selection \eqref{selection_ranks_average_rate} of the preferred serving cell rank indicator and the recommended interference rank indicator in the rank recommendation scheme. Unless explicitly mentioned, the cycling pattern over the transmission rank used in the Master-Slave scheduler is based on $I_1^{(i)}$,$I_2^{(i)}$,$I_1^{(i)}$,$I_2^{(i)}$,$I_3^{(i)}$ $\forall i$ and is determined only based on the number of rank recommendation requests. 

% Table 2
\begin{table*}%[!hhh]
    \centering
    \begin{threeparttable}
%    \captionstyle{center}
    \caption{System-level simulation assumptions.}
    \label{Table_sim_assumptions}
    \begin{tabular}{|c|cccc|}
    \hline
        \textbf{Parameter} & \multicolumn{4}{c|}{\textbf{Explanation/Assumption}} \\
    \hline \hline
     \multirow{4}{*}{Macro cell layout}
        & \multicolumn{4}{c|}{2-tier cellular system with wrap-around} \\
        & \multicolumn{4}{c|}{Hexagonal grid, 3-sector site (19 sites)} \\
        & \multicolumn{4}{c|}{Bore-sight points toward flat side} \\
        & \multicolumn{4}{c|}{10 users dropped per sector} \\
    \hline
        {Carrier frequency}
        & \multicolumn{4}{c|}{2 GHz} \\
    \hline
        {System bandwidth}
        & \multicolumn{4}{c|}{FDD: 10 MHz (downlink only)} \\
    \hline
        {Inter-site distance}
        & \multicolumn{4}{c|}{500 m} \\
    \hline
        Antenna configuration
        & \multicolumn{4}{c|}{$4 \times 4$ uniform linear array with 4 $\lambda$ spacing at BS and 0.5 $\lambda$ spacing at user terminal} \\
    \hline
        \multirow{3}{*}{Channel model} 
        & \multicolumn{4}{c|}{Spatial channel model}\\
        & \multicolumn{4}{c|}{Urban macro based on 3GPP case 1 with 3km/h mobility} \\
        & \multicolumn{4}{c|}{$15^{\circ}$ down-tilting and $15^{\circ}$ angle spread} \\    
    \hline
        Subband size & \multicolumn{4}{c|}{6 RB (subband)} \\
    \hline
        Scheduling
        & \multicolumn{4}{c|}{Proportional fair in time/frequency domains} \\
    \hline
        Resource allocation
        & \multicolumn{4}{c|}{RB-level indication} \\
    \hline
        \multirow{3}{*}{Transmission mode}
        & \multicolumn{4}{c|}{Single-user MIMO with and without rank coordination} \\
        & \multicolumn{4}{c|}{Triggering threshold $\delta$ in \eqref{measurement_set}: 10dB} \\
        & \multicolumn{4}{c|}{Inter-site clustering: 3 cells (sectors) per cluster} \\ 
    \hline
        Modulation and coding
        & \multicolumn{4}{c|}{MCS based on LTE transport formats} \\
    \hline
        Link abstraction & \multicolumn{4}{c|}{Mutual Information Effective SINR Mapping MIESM (Received Bit Mutual Information Rate RBIR)} \\
    \hline
        \multirow{2}{*}{Hybrid ARQ}
        & \multicolumn{4}{c|}{Chase combining, non-adaptive/asynchronous} \\
        & \multicolumn{4}{c|}{Maximum 3 retransmissions} \\
    \hline
        \multirow{6}{*}{Feedback}
        & \multicolumn{4}{c|}{RI (wideband): 2 bit} \\
        & \multicolumn{4}{c|}{Recommended interference rank (wideband): 2 bit} \\
        & \multicolumn{4}{c|}{PMI (wideband/subband): 4 bit LTE codebook} \\
        & \multicolumn{4}{c|}{CQI (wideband/subband): 4 bit CQI} \\
        & \multicolumn{4}{c|}{5 ms (period), 6 ms (delay)} \\
        & \multicolumn{4}{c|}{No feedback errors} \\
    \hline
        Channel estimation & \multicolumn{4}{c|}{ideal and non-ideal (mean-square error obtained from link level curves)} \\
    \hline
        \multirow{2}{*}{Link adaptation}
        & \multicolumn{4}{c|}{Target block error rate: 10 \%} \\
        & \multicolumn{4}{c|}{(ACK: +0.5/9 dB, NACK: -0.5 dB)} \\
    \hline
        Traffic model & \multicolumn{4}{c|}{Full buffer} \\
    \hline
        \multirow{2}{*}{Network}
        & \multicolumn{4}{c|}{Synchronized} \\
        & \multicolumn{4}{c|}{Fast backhaul}\\
    \hline
    \end{tabular}
    \end{threeparttable}
%\vspace{3cm}
\end{table*}

\par The performance is measured in terms of the average cell spectral efficiency (``Average throughput'') and the 5\% cell edge spectral efficiency (``cell-edge throughput'').

\begin{figure}%[!hhh]
\centerline{\includegraphics[width=0.7\columnwidth]{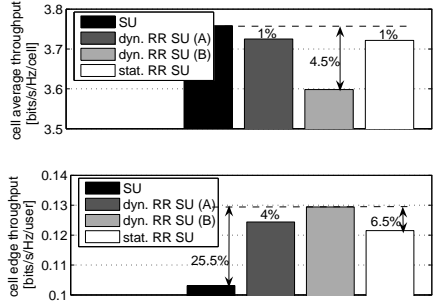}}
  \caption{Performance achievable by dynamic (dyn.\ RR SU) and statistical (stat.\ RR SU) rank coordination over single-cell SU-MIMO in a $n_t \times n_r=4\times 4$ ULA (4,15).}
  \label{RR_dynamic_statistical}
\end{figure}

\par Figure \ref{RR_dynamic_statistical} provides the performance achievable for a minimum mean square error (MMSE) receiver with ideal Iinterference Rejection Capability (IRC) that relies on an estimate of the interfering cell user-specific demodulation reference signals (DM-RS) to build the interference covariance matrix. 

\par We investigate the gain of coordination for various cycling patterns. With the dynamic cycling pattern $I_1^{(i)}$,$I_2^{(i)}$,$I_1^{(i)}$,$I_2^{(i)}$,$I_3^{(i)}$, denoted as (A) in Figure \ref{RR_dynamic_statistical}, we observe that a gain of 20.7\% is achieved at the cell edge by the proposed rank recommendation-based Master-Slave coordinated scheduling scheme over the baseline (without coordination) system with only 2 extra feedback bits! The slight loss at the cell average can be recovered by slightly tweaking the PF parameter. A second dynamic cycling pattern $I_1^{(i)}$,$I_2^{(i)}$,$I_1^{(i)}$,$I_2^{(i)}$,$I_1^{(i)}$, denoted as (B) in Figure \ref{RR_dynamic_statistical}, is also investigated where more stress is given to cell edge users as the last entry of the pattern has been switched to $I_1$. Contrary to the first pattern, the second pattern has a non-negligible cell average throughput loss because $I_1^{(i)}$ and $I_2^{(i)}$ are most of the time equal to 1 and 2 $\forall i$, and, therefore, users in the Master cell with the preferred RI equal to 3 and 4 have few chance to be scheduled. 
Recall that if $\mathcal{U}_{\textrm{M},1}\neq\emptyset$, BS M schedules only users belonging to $\mathcal{U}_{\textrm{M},1}$. It helps cell edge users because they have more chance to be scheduled and benefit from the rank recommendation. The cycling pattern $I_1^{(i)}$,$I_2^{(i)}$,$I_1^{(i)}$,$I_2^{(i)}$,$I_3^{(i)}$ outperforms $I_1^{(i)}$,$I_2^{(i)}$,$I_1^{(i)}$,$I_2^{(i)}$,$I_1^{(i)}$ in terms of cell average throughput because $\mathcal{U}_{\textrm{M},1}$ is often empty in the subframe whose transmission rank is fixed to $I_3^{(i)}$, therefore allowing Master BS to schedule rank 3 and 4 users frequently.

\par When an MMSE receiver with ideal IRC is used, the preferred interference rank is most of time equal to 1. Such statistical information can be used to reduce the feedback overhead and simplify the cycling mechanism in the scheduler. Indeed, rather than requesting the CoMP users to report the preferred interference rank and dynamically update the cycling pattern as in $I_1^{(i)}$,$I_2^{(i)}$,$I_1^{(i)}$,$I_2^{(i)}$,$I_3^{(i)}$, we can simply assume that the preferred interference rank of CoMP users is equal to 1 and operate the coordinated scheduler by pre-defining the cycling pattern. To that end, we also evaluate in Figure \ref{RR_dynamic_statistical} the case where the same cycling pattern 1,2,1,2,3 is fixed in all cells (denoted as stat.\ RR SU). The predefined cycling still enables to get a significant cell edge improvement of 18\%. Only a slight loss is observed compared to the case where the interference rank is reported and the cycling pattern is dynamically updated based on that report, as with $I_1^{(i)}$,$I_2^{(i)}$,$I_1^{(i)}$,$I_2^{(i)}$,$I_3^{(i)}$. Statistical rank recommendation has the advantage that multi-cell coordination can achieve a cell edge performance gain without increasing the feedback compared to a baseline system requiring no coordination. It still relies on messages exchanges between cells to achieve the coordination. Note that the pre-defined cycling pattern is receiver implementation specific, contrary to the dynamic cycling patterns (A) and (B).

\begin{figure}%[!hhh]
\centerline{\includegraphics[width=0.7\columnwidth]{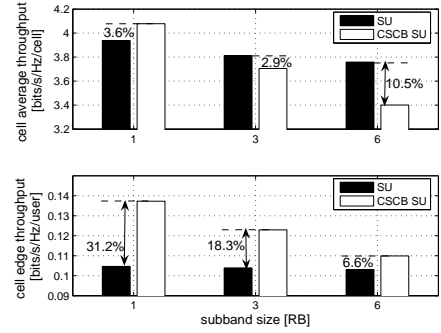}}
  \caption{Performance achievable by iterative CSCB (CSCB SU) over single-cell SU-MIMO (SU) in a $n_t \times n_r=4\times 4$ ULA (4,15).}
  \label{CSCB_SU}
\end{figure}

\par Figure \ref{CSCB_SU} evaluates the performance of a state-of-the-art iterative coordinated scheduling and beamforming (iterative CSCB) scheme relying on the signal to leakage and noise ratio (SLNR) criterion \cite{Sadek:2007} and the architecture introduced in \cite{Yu:2011}, as a function of the subband size. The power on each beam is assumed binary (ON-OFF) controlled. Coordination is performed at the whole network level with 57 cells (in contrast with the 3-cells clustering assumed for rank coordination) and the maximum number of inter-cell iterations before actual scheduling is fixed to 8. The feedback for the iterative CSCB (with a triggering threshold of 10dB) assumes unquantized explicit feedback (contrary to the quantized implicit feedback assumed in rank coordination) with the average channel matrices reported per 1RB, 3RB and 6RB subband. The performance gain of CSCB with accurate feedback (1RB) provides significant gain (31\%) over uncoordinated SU-MIMO. However, even with unquantized feedback and a large number of cooperating cells, the performance of the iterative CSCB drops significantly as the subband size increases. The BS has to compute the CQI, beamformers and transmission rank at every iteration after performing interference suppression and multi-cell coordination. However, given the high frequency selectivity of the spatially uncorrelated channel and the feedback inaccuracy at the subband level, it is very complicated to accurately predict those quantities while accounting for cooperation (and explains for the big loss incurred by going from 1RB to 3 RB and to 6RB). The inaccurate CQI prediction hampers the appropriate selection of the user, the transmission ranks and the beamformers at every iteration of the scheduler and ultimately the whole link adaptation and convergence of the scheduler. Similar observations were made in \cite{Clerckx:2010} for SU and MU-MIMO but the effect is more pronounced for multi-cell cooperation. Most of the theoretical performance gain can therefore be lost because of the inaccurate link adaptation. It is worth noting that the receiver implementation (MMSE with ideal IRC) was assumed known at the BS and the feedback is unquantized in the iterative CSCB evaluations. The results presented here are therefore upper bound on the throughput achievable by the iterative CSCB in a more practical setup. 
\par Recalling that performance in Figure \ref{RR_dynamic_statistical} assumes 6RB subband size, by comparing Figures \ref{RR_dynamic_statistical} and \ref{CSCB_SU}, it is observed that the rank coordination shows very competitive performance compared to the iterative CSCB, with a lower feedback overhead and scheduler complexity. In rank coordination, the user computes the CQI accounting for the effect of coordination and the scheduler satisfies the user requests, therefore enabling a more accurate and simpler link adaptation than with the iterative CSCB.

\begin{figure}%[!hhh]
\centerline{\includegraphics[width=0.7\columnwidth]{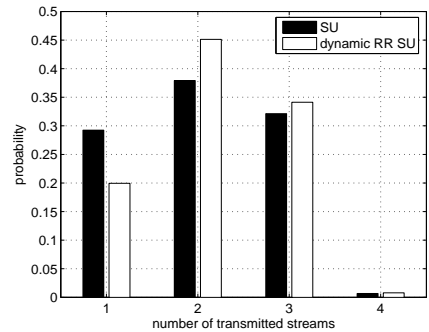}}
  \caption{Statistics of the transmission rank, i.e.\ the number of transmitted streams, with dynamic rank coordination and without rank coordination (single-cell SU-MIMO) in a $n_t \times n_r=4\times 4$ ULA (4,15).}
  \label{RR_rank}
\end{figure}

\par Figure \ref{RR_rank} shows the distribution of the actual transmission rank after scheduling for the baseline system without coordination and the rank recommendation-based coordinated scheduling when the dynamic cycling pattern $I_1^{(i)}$,$I_2^{(i)}$,$I_1^{(i)}$,$I_2^{(i)}$,$I_3^{(i)}$ and an MMSE receiver with ideal IRC are used. A large portion of users who used to be scheduled in rank 1 transmission in the baseline system benefit from rank 2 transmission in the rank recommendation-based coordinated scheduling scheme. It confirms that the joint selection of the preferred serving cell rank indicator and the preferred interference rank indicator combined with the Master-Slave scheduler enables higher rank transmissions even to cell edge users. 

\begin{figure}%[!hhh]
\centerline{\includegraphics[width=0.7\columnwidth]{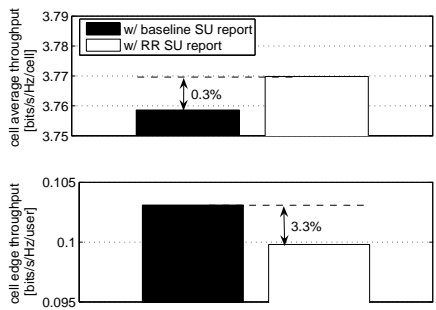}}
  \caption{Performance of the single-cell scheduler with baseline SU-MIMO report and rank recommendation-based report in a $n_t \times n_r=4\times 4$ ULA (4,15).}
  \label{RR_single_cell_scheduler}
\end{figure}

\par Figure \ref{RR_single_cell_scheduler} has a double objective: 1) illustrate the sensitivity of the algorithm to a mismatch between the assumptions on transmit precoding and base station coordination made by the UE at the time of CSI computation and the actual decisions of the scheduler, 2) illustrate the importance of combining the joint selection of the preferred serving cell rank indicator and the preferred interference rank with the Master-Slave coordinated scheduler to harvest cell-edge performance gains. The dynamic cycling pattern $I_1^{(i)}$,$I_2^{(i)}$,$I_1^{(i)}$,$I_2^{(i)}$,$I_3^{(i)}$ and an MMSE receiver with ideal ICI rejection capability are used. Intuitively, if the user reports the rank recommendation-based feedback information but the scheduler relies on a baseline (without coordination) scheduler, performance may be impacted as the reported preferred serving cell rank $R_{q}^{\star}$ can be over-estimated and the assumptions made about coordination by the UE are not followed by the base stations. To assess that impact, we investigate the performance of a single-cell (denoted as baseline) scheduler when two different feedback information are reported: the reported preferred serving cell rank and CQI as the ones computed in the baseline system and as the ones computed assuming rank recommendation. As we can see from Figure \ref{RR_single_cell_scheduler}, no gain (or even a slight loss at the cell edge) is observed because of the lack of appropriate coordination. 

\begin{figure}%[!hhh]
\centerline{\includegraphics[width=0.7\columnwidth]{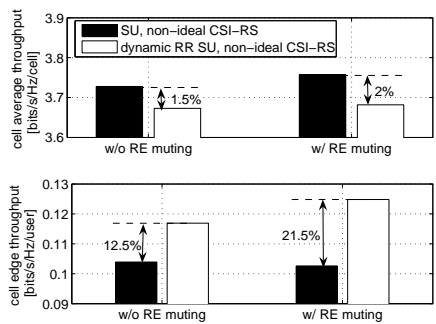}}
  \caption{Performance gain of the rank recommendation over single-cell SU-MIMO with and without RE muting in the presence of CSI-RS channel estimation errors.}
  \label{RR_channel_estimation}
\end{figure}

\par Figure \ref{RR_channel_estimation} evaluates the performance of rank coordination in the presence of estimation errors on the reference signals used for channel measurement (denoted as CSI-RS in LTE-A).  The mean square channel estimation error as a function of the wideband SINR is first computed based on a link level simulator and is applied to the system level simulator. From Figure \ref{RR_channel_estimation}, we note that multi-cell coordination is affected by the CSI-RS measurement errors even though the recommended rank is a wideband information. Despite this sensitivity, a 12.5\% gain at the cell edge is still achievable compared to a network not relying on multi-cell coordination. In order to recover the loss generated by CSI-RS measurement errors, we perform resource muting (as standardized in LTE-A) in the adjacent cell and evaluate the performance of the rank coordination in the presence of CSI-RS measurement errors. The resource muting coordination between cells allows for a better reception of CSI-RS of the other cells and at the same time better channel measurement accuracy for the CSI-RS of the serving cell. With resource muting, the rank coordination is shown to recover most of the gain achievable with perfect channel estimation.

\begin{figure}%[!hhh]
\centerline{\includegraphics[width=0.7\columnwidth]{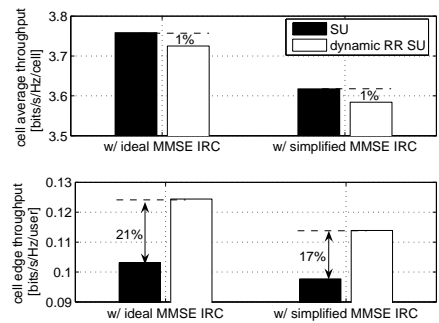}}
  \caption{Performance gain of the rank recommendation over single-cell SU-MIMO with ideal and simplified MMSE IRC (interference rejection combining) receiver.}
  \label{RR_receiver}
\end{figure}

\par Figure \ref{RR_receiver} illustrates that the coordination scheme provides significant gains also with other types of receivers, namely a MMSE receiver with a simplified ICI rejection capability (not relying on the DM-RS measurement of the interfering cells). It computes the receiver filter using an estimate of the covariance matrix of the interference by assuming the precoder in the interference cells is the identity matrix. We also observe a significant gain of roughly 17\% at the cell edge with the proposed rank recommendation-based Master-Slave coordinated scheduling scheme over the baseline (without coordination) system.

\section{Conclusions}
We introduce a novel and practical interference mitigation technique relying on a dynamic coordination of the transmission ranks among cells in order to help cell edge users to benefit from higher rank transmissions. The coordination requires the report from the users of a recommended rank to the interfering cells. Upon reception of those information, the interfering cells coordinate with each other to take informed decisions on the transmission ranks that would be the most beneficial to the victim users in neighboring cells and maximize a network utility function. Such method is shown to provide significant cell-edge performance gain over uncoordinated LTE-A system under a very limited feedback and backhaul overhead. It enables efficient link adaptation and is robust to channel measurement errors.

\begin{biography}[{\includegraphics[width=1in,height=1.25in,clip,keepaspectratio]
{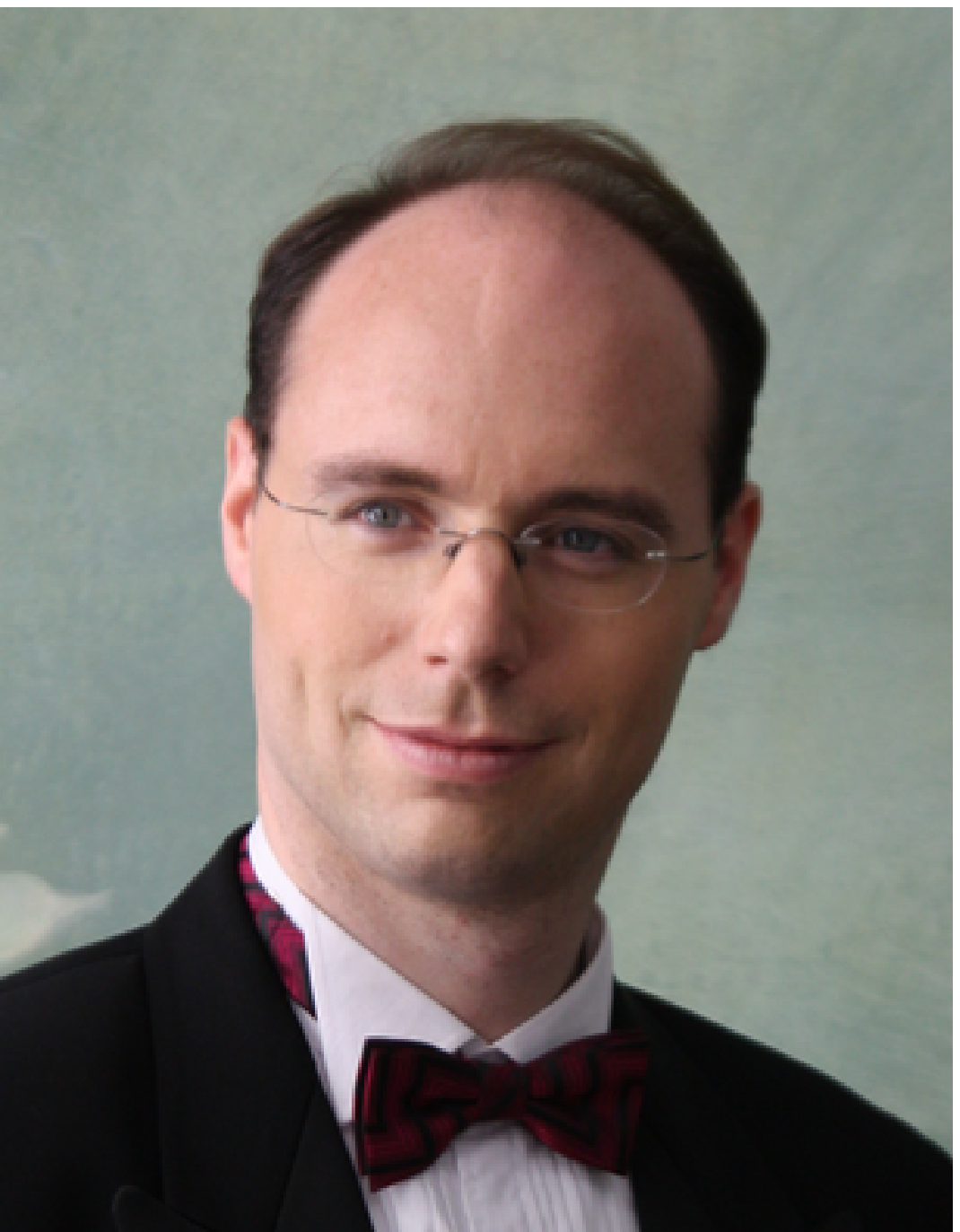}}]{Bruno Clerckx} received his M.S. and Ph.D. degree
in applied science from the Universite catholique
de Louvain (Louvain-la-Neuve, Belgium) in 2000
and 2005, respectively. He held visiting research
positions at Stanford University (CA, USA) in 2003
and Eurecom Institute (Sophia-Antipolis, France) in
2004. In 2006, he was a Post-Doc at the Universite
catholique de Louvain. From 2006 to 2011, he was
with Samsung Electronics (Suwon, South Korea)
where he actively contributed to 3GPP LTE/LTE-A
and IEEE 802.16m and acted as the rapporteur for
the 3GPP Coordinated Multi-Point (CoMP) Study Item and the editor of the
technical report 3GPP TR36.819. He is now a Lecturer (Assistant Professor)
in the Electrical and Electronic Engineering Department at Imperial College
London (London, United Kingdom).
He is the author or coauthor of two books on MIMO wireless communications
and numerous research papers, standard contributions and patents. He
received the Best Student Paper Award at the IEEE SCVT 2002 and several
Awards from Samsung in recognition of special achievements. Dr. Clerckx
serves as an editor for IEEE TRANSACTIONS ON COMMUNICATIONS. 
\end{biography}

\begin{biography}[{\includegraphics[width=1in,height=1.25in,clip,keepaspectratio]
{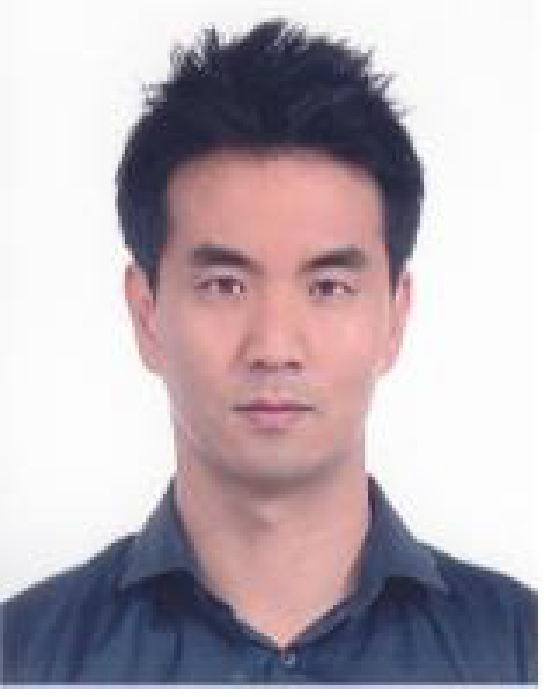}}]{Heunchul Lee} received the B.S., M.S., and Ph.D. degrees in electrical
 engineering from Korea University, Seoul, Korea, in 2003, 2005, and
 2008, respectively. From February 2008 to October 2008 he was a
 Post-doctoral Fellow under the Brain Korea 21 Program at the same
 university. From November 2008 to November 2009 he was a Post-doctoral
 Fellow in information systems Laboratory at Stanford University under
 supervision of Professor A. Paulraj. Since January 2010 he has been
 with Samsung Electronics, where he is a senior engineer, currently,
 working in LTE/LTE-A Modem design. During the winter of 2006, he
 worked as an intern at Beceem Communications, Santa Clara, CA, USA.
 His research interests are in communication theory and signal
 processing for wireless communications, including MIMO-OFDM systems,
 multi-user MIMO wireless networks, Wireless Body-area networks (WBAN)
 and 3GPP LTE/LTE-A. Dr. Lee received the Best Paper Award at the 12th
 Asia-Pacific conference on Communications, and the IEEE Seoul Section
 Student Paper Contest award, both in 2006. In addition, he was awarded
 the Bronze Prize in the 2007 Samsung Humantech Paper Contest in
 February 2008.
\end{biography}
 
\begin{biography}[{\includegraphics[width=1in,height=1.25in,clip,keepaspectratio]
{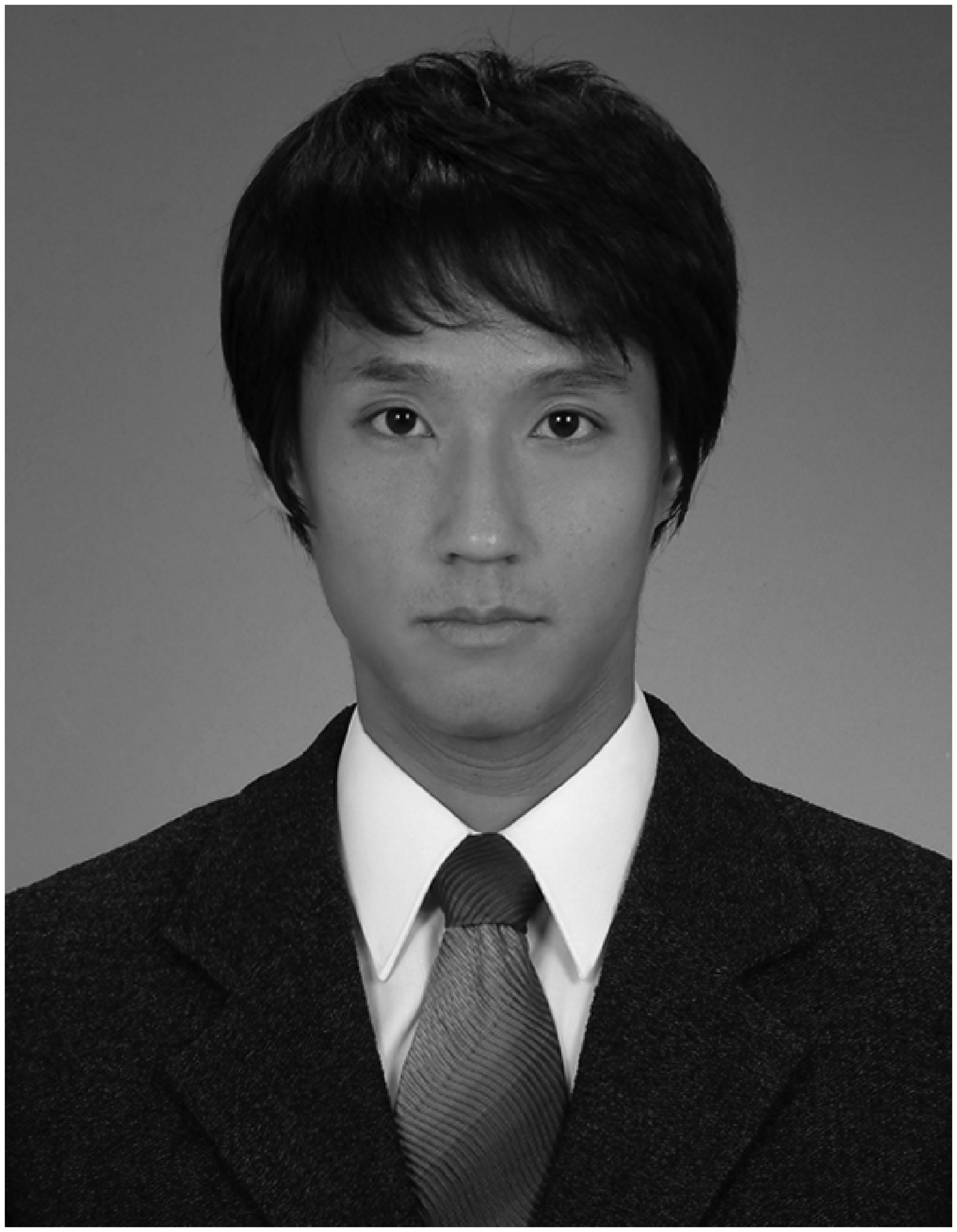}}]{Young-Jun Hong} (S'04--AM'09)
received the B.S. degree in electrical engineering from Yonsei University, Seoul, Korea, in 2000, and the M.S. and Ph.D. degrees in electrical engineering from the Korea Advanced Institute of Science and Technology (KAIST), Daejeon, Republic of Korea, in 2002 and 2009, respectively.
He is currently a Senior Engineer with Samsung Electronics Co., Ltd., Korea, since 2009.
He served as a delegate in 3GPP Long Term Evolution-Advanced (LTE-A) and his standard activities with Samsung Electronics included coordinated multi-point (CoMP) transmission and reception, coordinated scheduling/coordinated beamforming (CS/CB), inter-cell interference coordination (ICIC), and heterogeneous network (HetNet).
He is currently working on the research and development of ultra-low power mixed-signal and digital integrated circuit and wireless wearable sensor platform in the area of Medical Body Area Network (MBAN).
\end{biography}

\begin{biography}[{\includegraphics[width=1in,height=1.25in,clip,keepaspectratio]
{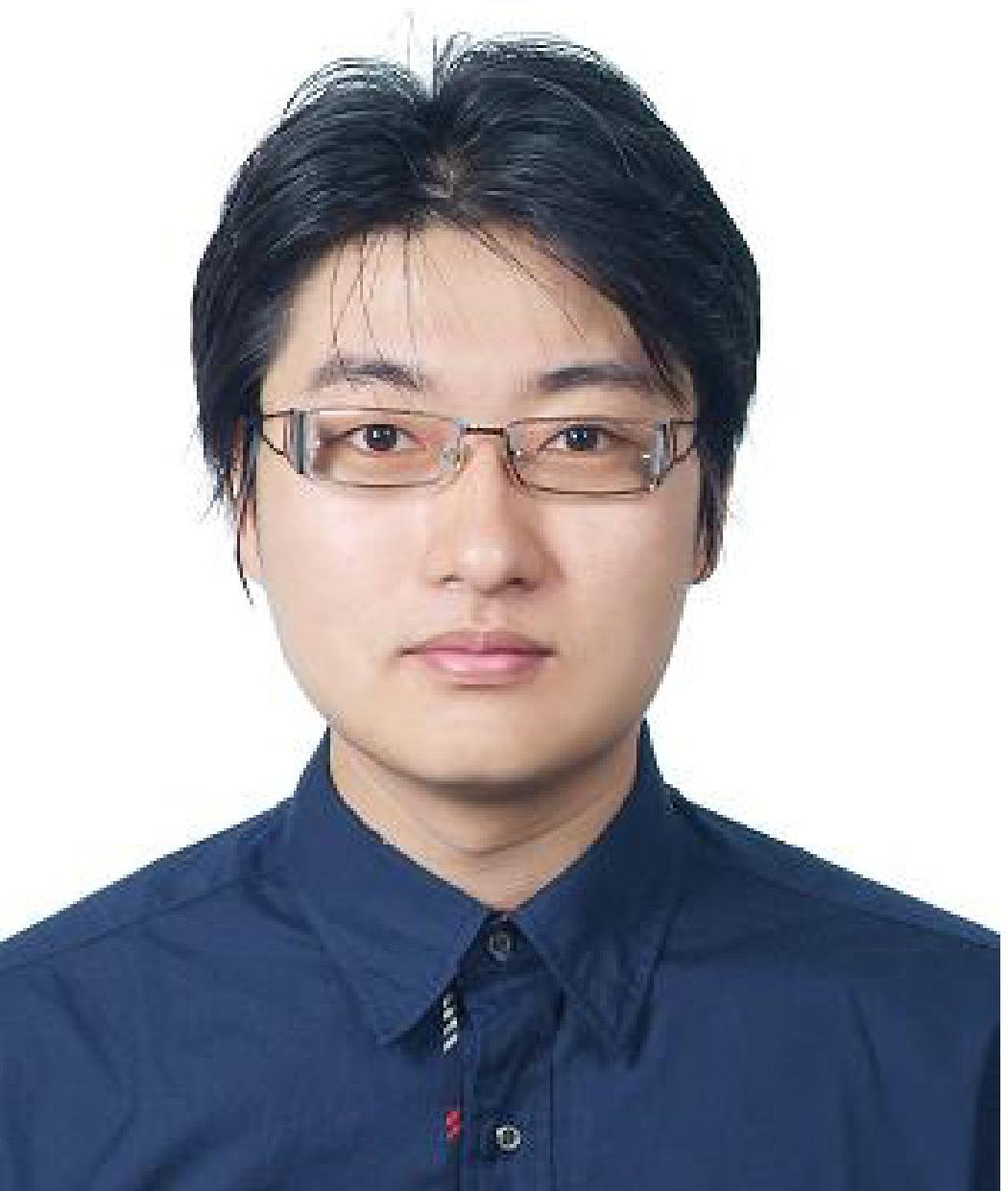}}]{Gil Kim} received his B.S. and M.S. degrees from Seoul National University in Seoul, South Korea in 2005 and 2007, respectively. From 2007 to 2010, he was with Samsung Advanced Institute of Technology (Giheung, South Korea) as a member of R\&D staff. He were engaged in the design and standardization of physical layer of 3GPP LTE-Adv systems. He is now a senior engineer in Samsung Electronics (Suwon, South Korea) where he is primarily involved in the system design of 4G modems for terminals. His research area includes MIMO, cooperative systems, and beyond 4G wireless communications.
\end{biography}

\end{document}